# The nucleon phase hypothesis of binary fission

G. Mouze, S. Hachem, and C. Ythier
Faculté des Sciences, Université de Nice, 06 108 Nice cedex 2, France.

The mass distribution of fission fragments of actinide and superheavy nuclei can be explained if a new state of nuclear matter, a nucleon phase, is created in any fission event.

## 1. Introduction.

If a nuclear fission can be used for making nuclear explosives, it may be asked whether the nuclear fission process is not already, per se, a kind of explosion.

It has been recently shown [1] that the rearrangement of fissioning systems such as $^{233}U + n_{th}$, $^{235}U + n_{th}$, $^{239}Pu + n_{th}$ and $^{252}Cf$ (s.f.) into two nascent light and heavy fragments occurs within 0.17 yoctosecond.

We show, in Sect.2, that extreme conditions of energy and temperature are created in this rearrangement. We suggest, in Sect.3, that the proton phase and the neutron phase, which coexist in nuclear matter, could then be changed into a unique phase, a "nucleon phase", in which any distinction between proton and neutron has been abolished, but in which "nucleons" form closed shells of 82 and 126 nucleons. We further show that this hypothesis explains the mass distributions of asymmetric fission (Sect.4), and even those of symmetric fission (Sect.5).

## 2. The extreme conditions of nuclear fission.

During the extremely short time of 0.17 yoctosecond of the rearrangement process, a considerable amount of energy becomes available in the fissioning system. The energy-time uncertainty relation allows its determination:

$$\Delta E = \hbar / \Delta t = 6.582118\ 10^{-16}\ \text{eV s} / 1.70\ 10^{-25}\ \text{s} = 3.86\ \text{GeV}, \qquad (1)$$

This energy is considerably greater than the energy of the most energy-rich fragment pair of the $^{235}U + n_{th}$ –system, which is equal to only 205.88 MeV [2].

This energy is related, by the Boltzmann constant $k = 8.6173\ 10^{-5}$ eV $K^{-1}$, to a temperature T of the order of

T (Kelvin) = $\Delta E / k = \sim 4.5\ 10^{13}$ K.

And such estimation is useful to measure the temperature of nuclear explosions.

It may be asked whether, under so extreme conditions, a new state of nuclear matter could not be created. The answer could be found in Terrell's work on prompt neutron emission.

## 3. The nucleon-phase hypothesis

### 3-1 Terrell's work on prompt neutron emission

H.H. Knitter et al. [3] report that "the first meaningful systematization of neutron multiplicities is from Terrell], *"who demonstrated that several $\bar{\nu}$ (A) curves almost coincide, and suggested that fragment rather than compound nucleus properties should fix the multiplicities ".*

Indeed, Terrell [4] has plotted the data concerning $^{233}U + n_{th}$, $^{235}U + n_{th}$, $^{239}Pu + n_{th}$ and $^{258}Cf$ (s.f.) in a single graph (fig.11 of ref.4), and shown that "the results for neutron emission from these four different fissioning systems are so strikingly similar, that the neutron yield as a function of mass could be represented rather accurately by a single curve, made of straight lines in the light and heavy fragment regions": consequently, the mean value of the neutron yield could be given by their sum

$$\bar{\nu} = 0.08\ (A_L - 82) + 0.10\ (A_H - 126). \qquad (3)$$

It is the famous equation of Terrell, which remained unexplained up to now.

Moreover, Terrell reported the important observation that asymmetric fission seems to be characterized by the relations:

$$A_L > 82, \quad A_H > 128. \qquad (4)$$

He wrote: "these limits seem to define quite accurately the regions of appreciable yield for asymmetric fission. They also seem to be the points at which neutron yield nearly vanishes".

As for the use of 126 instead of 128 in eq. (3), he wrote: "the mass number 126 is used here instead of 128 because it gives a better linear representation".

These words suggest that *magic mass numbers* 82 and 126 might exist, and play a role in the emission of prompt neutrons and in the limits of the regions of appreciable mass yield of fission fragments.

Only a sophisticated critical analysis of the experimental data on prompt neutron emission and the constant recourse to the mathematical methods of statistics could lead Terrell to so important conclusions, e.g. that there is no neutron emission at symmetry. Let us recall that he had determined, in 1957 [5], the value $\sigma$ = 1.08 of the parameter of the Gaussian curve representing the probability $P(\nu)$ of emitting $\nu$ neutrons per fission -- at that time believed to be the standard deviation characterizing "the fragment excitation energy " [1]---.

**3-2 The new hypothesis**

At the International Winter Meeting on Nuclear Physics held in Bormio in January 2008, G. Mouze and C. Ythier announced that Terrell's equation could be interpreted as revealing the role of an intermediary "nucleon phase" in asymmetric fission [6].

They asserted that Terrell's equation could suggest that the prompt neutrons are emitted by the valence shells of an "A = 82 nucleon core" and by the valence shells of "an A = 126 nucleon core", whereas the A = 82 and A =126 cores themselves do not emit any neutron, *as if a nucleon phase could be created, in which "nucleon shells" are closed at A = 82* in the nascent light fragment*, and at A* = 126 in the nascent heavy fragment.

More precisely, Terrell's equation suggests that the number of emitted prompt neutrons is proportional to the nucleon density in the valence shells of these new "nucleon cores": This number increases with the mass of the shells, is greater in the heavier shell, and is zero for an empty shell.

Before we show how the nucleon phase hypothesis explains the *asymmetric* distribution of the light actinide nuclei and the *symmetric* mass distributions of $^{258}Fm$ and superheavy nuclei, let us ask ourselves how such a nucleon phase can suddenly be created within a fissioning system. In other words, what is the necessary "ignition step" of the extreme conditions of the rearrangement reaction?

**3-3 The "ignition" step of the nucleon phase.**

At the 1988 Karlsruhe Symposium on Transuranium Elements Today and Tomorrow the following idea of G. Mouze was presented [7]: the reaction

$^{239}Pu + n \rightarrow\ ^{132}Sn +\ ^{108}Ru + 216.87$ MeV has to be considered as a two-step process, in which the first step is an internal rearrangement, now called clusterisation,

$$^{239}Pu + n \rightarrow\ ^{208}Pb +\ ^{32}Mg + 81.3 \text{ MeV}. \quad (5)$$

This energy- yielding process could be expected, because it looks like the reverse process of an energy-requiring Oganessian reaction, such as:

$^{208}Pb$ + 291- MeV $^{58}Fe \rightarrow {}^{265}Hs + n$ [8]. (6)

It was also pointed out that the energy released by this clusterization could be communicated to the subshells of $^{208}Pb$ , because it could be changed into vibrational energy and lead to *an internal core-cluster collision*. And if all the 76 valence nucleons belonging to the deep-lying $^{132}Sn$ core make a transition to the free states of the $^{32}Mg$ cluster, the light fragment $^{108}Ru$ could be synthesized in the second step:

$^{208}Pb + {}^{32}Mg \rightarrow {}^{132}Sn + {}^{108}Ru$ + 135.5 MeV.

In the nucleon- phase model, the maximum number of transferred nucleons is no more 76, but 82. The reason is that an *A = 126 nucleon core is created in the $^{208}Pb$-core, as soon as this $^{208}Pb$ core collides with its cluster*, and further that the nucleons surrounding this A = 126 core behave as valence nucleons, which can be transferred to the primordial cluster: their number is 208 - 126 = 82.

## 4. Justification of the limits of the mass yield curves

**4-1 The asymmetric case.** There, the greatest value of the mass number $A_L$ of the light fragment is given by

$$A_L^{MAX} = A_{cl} + \underline{82}, \qquad (8)$$

since *82* nucleons can be transferred from the A = 126 nucleon core to the cluster, of mass $A_{cl}$; and the smallest value of $A_L$ is

$$A_L^{Min} = 82 , \qquad (9)$$

since an A = 82 nucleon core has been formed around the cluster in the nascent light fragment . Similarly

$$A_H^{MAX} = 126 + [\underline{82} - (82 - A_{cl})] = 126 + A_{cl} , \qquad (10)$$

$$A_H^{MIN} = 126 . \qquad (11)$$

One sees that *the width ΔA* of the region of appreciable yield, for the light fragment and for the heavy fragment, *is equal to $A_{cl}$ :*

$$\Delta A = A^{MAX} - A^{MIN} = A_{cl} \qquad (\textit{asymmetric fission}). \qquad (12)$$

This *rule,* demonstrated in ref. [9], holds for all actinide nuclei up to $^{252}Cf$.

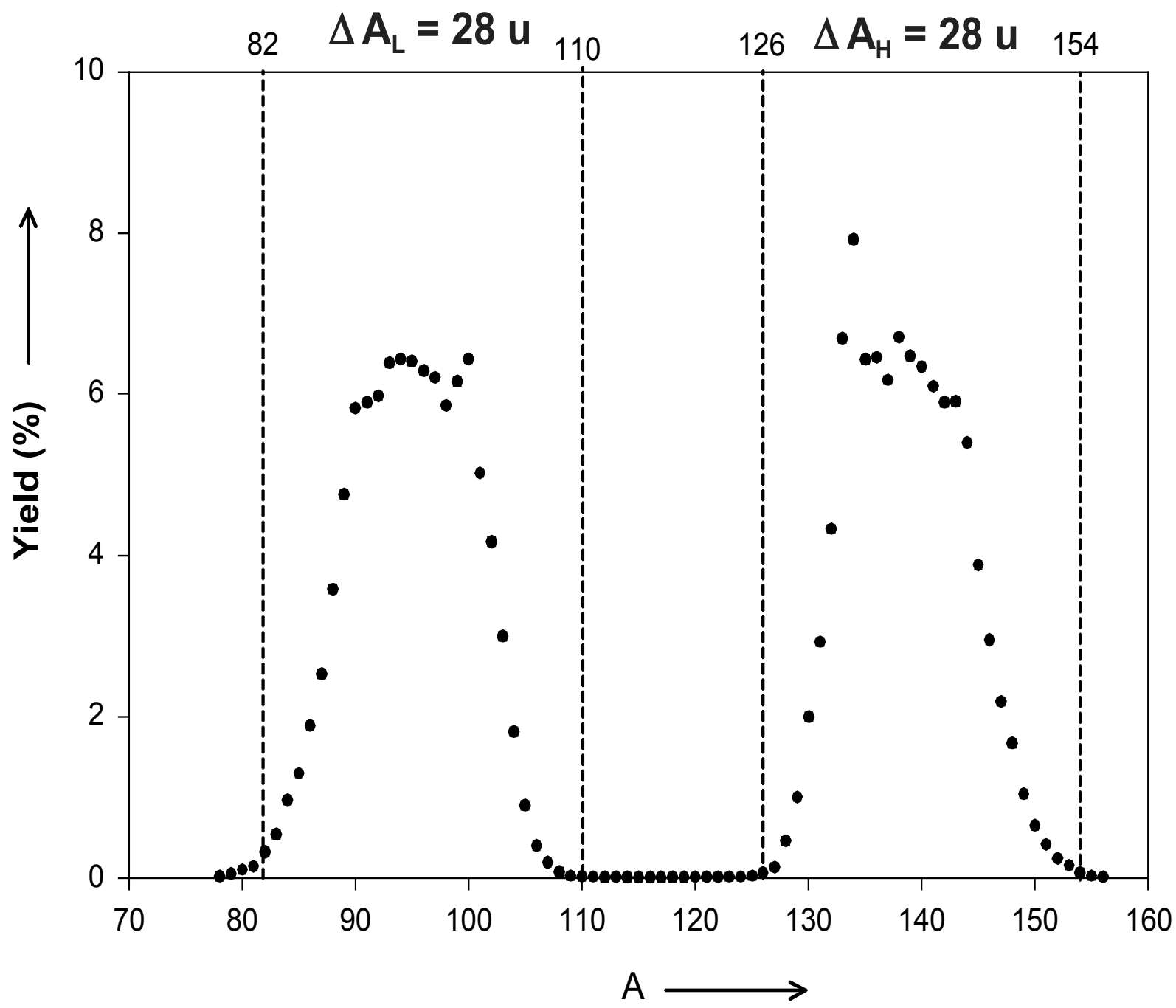

**Fig.1**: Influence of the nucleon phase on the mass yield of $^{235}U + n_{th}$:
The nucleon phase decides on the width of the region of appreciable yield of the fragments; this width is equal, in mass units, to the mass number $A_{cl}$ of the primordial cluster, here $^{28}Ne$. Note that no correction has been made for the emission of prompt neutrons: the yields, taken from Flynn and Glendenin [10] also are those of the fission products. The decrease of the yield in the vicinity of the limits demonstrates that asymmetric fission is "confined" by the Coulomb barrier of the fragment pairs.

It is noteworthy that the mass distributions of $^{233}U + n_{th}$, $^{235}U + n_{th}$ and $^{239}Pu + n_{th}$ almost coincide precisely at A = 82, and that the mass yield of the light fragments decreases abruptly as soon as $A_L$ becomes smaller than 82, as noted by Terrell. Indeed, at $A_L = 82$ the yield is equal to $10^{-2}$ for these fissioning systems; but the yield becomes, for example for $^{235}U + n_{th}$, as small as $10^{-6}$ at $A_L = 71$. This behavior constitutes a strong argument in favor of the nucleon-phase hypothesis, because it is exactly what is expected in a nucleon phase, if closure of shells at magic numbers is a universal organization law of nuclear matter, which holds as well in a "nucleon" phase, for *magic mass numbers*, as it holds in the usual "proton" and "neutron" phase, for magic Z and N numbers.

Fig.1 shows the regions of appreciable yield of $^{235}U + n_{th}$; they extend from $A_L = 82$ to $A_L = 110$, and from $A_H = 126$ to $A_H = 154$; there ΔA= 28 u, in accordance with the rule (12), because the primordial cluster is $^{28}Ne$. As another example, the regions of appreciable yield for the system $^{239}Pu + n_{th}$ (fig.2) extend from $A_L = 82$ to $A_L = 114$, and from $A_H$ = 126 to $A_H$ =158. The region of appreciable yield of $A_L$ of all actinide nuclei up to $^{252}$Cf can be represented by the hatched area in the left part of fig.2.

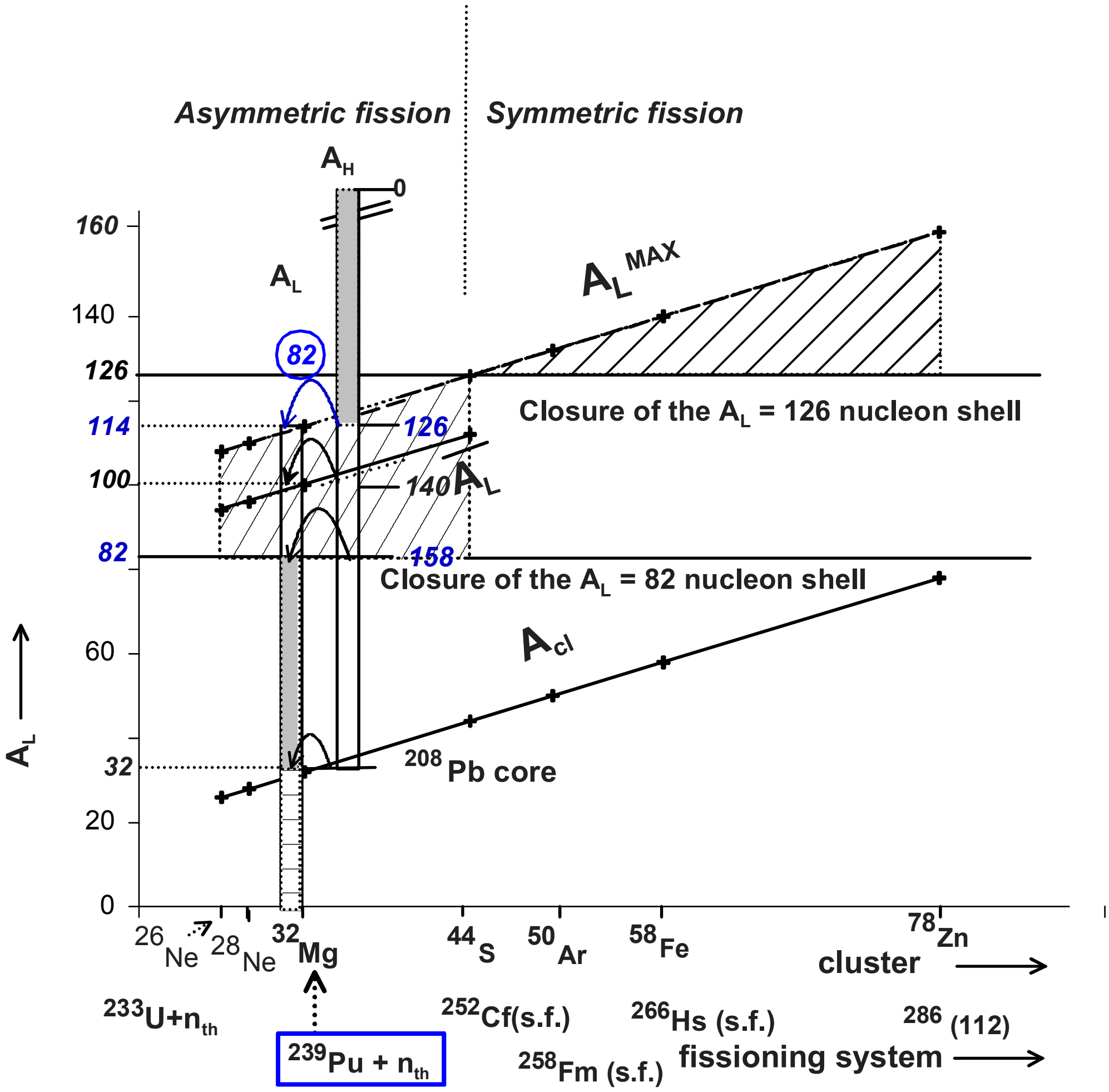

**Fig.2**: Transfer of 82 nucleons from core to cluster in the fission of $^{239}$Pu + $n_{th}$ according to the nucleon-phase model.
The region of appreciable yield for the asymmetric fission of the light actinide nuclei is represented by the hatched area on the left.

It is further noteworthy that *asymmetric fission*, characterized by an almost-zero yield at symmetry, *is a consequence of the laws of the nucleon phase*. Indeed, for all actinide nuclei lighter than $^{252}$Cf, it is not possible to form a single fragment having a mass $A_L$ equal to $A_F/2$. For example for $^{235}$U + $n_{th}$, $A_F/2$ = 118, whereas $A_L^{MAX}$ = $A_{cl}$ + 82 = 110. Even for $^{252}$Cf, only one light fragment of the region of appreciable yield can have a mass equal to $A_F/2$= 126, it is $A_L^{MAX}$ = 44 + 82 = 126; but the region of appreciable yield still extends from 82 to 126 and has a width Δ A =

44 u, equal to the mass number of the primordial cluster $^{44}S$, as for all asymmetric fissions.

**4-2 The new expression of the law of Flynn et al.**

The linear law of variation of the mean mass of the light fission product as a function of the mass $A_F$ of the fissioning system, first formulated by Flynn et al.[11], and later considered to be valid up to mendelevium nuclei [12], was interpreted by Mouze and Ythier [13] as the linear law of variation of this mean mass *as a function of the mass number* $A_{cl}$ of the primordial cluster, since $A_F$ can be written as A ($^{208}Pb$ + cluster). And it could be shown that the mean number of nucleons transferred in the synthesis of the light fission product is of the order of 66.7 [14] for $^{235}U + n_{th}$.

According to the nucleon-phase hypothesis, the mean value of the light fragment mass, $\overline{A_L}$, can now be written:

$\overline{A_L} = A_{cl} + \sim 68$ (*new linear law of variation*), (13)

whereas the mean value of the heavy fragment mass, $\overline{A_H}$, remains constant and equal to

$\overline{A_H} = \sim 140.$ (14)

Indeed, the mean mass of the heavy fission *product* has been found constant (see, e.g., [15]); for fissions induced by thermal neutrons or by reactor neutrons, this mass is equal to ~ 138 u [10]. In absence of excitation of the nucleus, the mean number of emitted prompt neutrons is $\bar{\nu}$ = = 2.3 for $^{240}Pu$, and $\bar{\nu}$= 2.0 for $^{238}U$(s.f.) [16].Thus the mean mass of the heavy *fragment* is almost constant and equal to about 140 u for the light actinide nuclei. This justifie eq.(14).

Consequently, it may be considered that about 140 – 126 = 14 nucleons remain, on an average, on the A = 126 nucleon core of the nascent heavy fragment. Since the total number of transferrable valence nucleons is equal to 82, it means that only 82 – 14 = 68 nucleons are, on an average, transferred to the primordial cluster. This justifies eq.(13).

In fig.2, the variation of $\overline{A_L}$ as a function of $A_F$, or of $A_{cl}$, is represented by a straight line parallel to the variation of $A_{cl}$.

After a necessary correction for the prompt neutron emission, it could be seen from fig.1 that $\overline{A_L}$, equal to 28 + 68 = 96, coincides with the mass number, A* = 82 + $\Delta A_L/2$ = 96, of the middle of the region of appreciable yield for this ($^{235}U + n_{th}$)- system. But this situation is exceptional. For the system $^{233}U + n_{th}$, $\overline{A_L}$ is equal to 94 and smaller than A*, equal to 95, whereas, for the system $^{252}Cf$ (s.f.), $\overline{A_L}$ is equal to 112, and much greater than A*, equal to 104. Anyway, the profile of the mass distribution of the light and heavy fission fragments, in the vicinity of the limits $A_{MIN}$ and $A_{MAX}$, for the light actinides up to $^{252}Cf$, is essentially determined by the tunnel effect, as justified in Sect.5-2; this remark holds, in particular, for the fission products of $^{235}U + n_{th}$ shown in fig.1.

**4-3 Nucleon phase and chemical thermodynamics**

Equations (13) and (14) mean that, in the rearrangement of the primordial dinuclear system of asymmetrically fissioning nuclei, a constant number of nucleons , 82, are *distributed* between the valence shells of the $A_H$ = 126 nucleon core and the free states of the cluster in such a way that the ratio:

(Mean number of nucleons remained on the $A_H$ =126 core) / (mean number really transferred to the cluster) remains

constant and equal to C = 14/68 = 0.206. (15)

At first sight, this situation is similar to that encountered in the study of the distribution of one and the same body between two unmiscible solvents. For such a distribution, Nernst found in 1891 the following law *at equilibrium*:

(concentration in solvent1) / (concentration in solvent 2)= Constant . (16)

There, the variation $d\mu_1$ and $d\ \mu_2$ of the chemical potential of the body in the two solvents must be equal

$$d\mu_1 = d\ \mu_2, \qquad (17)$$

and this justifies eq.(16).

This observation suggests that the sharing out of the nucleons between core and cluster could be described in the framework of chemical thermodynamics.

**4-4. The $^{252}$Cf case**.

In $^{252}Cf$, the A = 126 nucleon shell becomes filled, because 82 valence nucleons have been transferred from the $A_H$ = 126 core to the primordial cluster $^{44}S$. And for the first time the regions of appreciable yield become adjacent: in principle, they extend from $A_L$ = 82 to $A_L$ = 126 and from $A_H$ = 126 to $A_H$ = 170.

But two anomalies can be observed in this asymmetrically fissioning nucleus: 1°) instead of coinciding with the mass distributions of the n-induced fission of $^{233}U$, $^{235}U$ and $^{239}Pu$ at A = 82, the mass distribution of $^{252}Cf$ (s.f.) is characterized by a smaller yield at this A-value; 2°) the threshold of the prompt neutron emission is not exactly $A_H$ = 126, as for these three fissioning systems, but rather at $A_H$ = 132 [17].
Can these slight anomalies be explained by the fact that a complete $A_L$ = 126 core is now formed in the light fragment and can compete with the $A_L$ = 82 nucleon core in the sharing out of the valence nucleons?

But the surprising property of $^{252}Cf$ is not this slight shift of the regions of appreciable yield towards greater A-values, it is rather that their width is still equal to $A_{cl}$ = 44, *whereas addition of only a few nucleons* will put an end to the asymmetric fission mode, as will now be shown.

In conclusion of this Sect.4, it may be asserted that the *nucleon phase hypothesis explains asymmetric fission*, since it explains the asymmetric mass distributions down to the last detail.

**5. The two origins of symmetric fission**

**5-1 The limits of the mass yield curves**

As soon as an A = 126 nucleon core can exist in a nucleus heavier than $^{252}Cf$, a new situation is created and eq.(9) is replaced by

$A_L^{MIN} = 126,$ (18)

while eq.(8), $A_L^{MAX} = A_{cl} + \underline{82}$, remains valid. Consequently, for a nucleus such as $^{258}Fm$, with $A_L^{MAX} = 50 + \underline{82} = 132$, the region of appreciable yield becomes much narrower than in the asymmetric case.

$\Delta A_L = A_L^{MAX} - A_L^{MIN} = 132 - 126 = 6$ u, (19)

and coincides with *symmetry* ( A* = 258/2 = 129). We may conclude:

*It is the closure of the $A_L$ = 126 nucleon shell in the nascent light fragment that causes the symmetric mass distribution.*

Moreover*, light and heavy fragments have now the same region of appreciable yield*, since $A_H^{MAX} = 126 + [\underline{82} - (126 - A_{cl})] = A_{cl} + \underline{82}$, also 132 for $^{258}Fm$, and $A_H^{MIN} = 126$.

Their common width of appreciable yield is now given by

$\Delta A = (A_{cl} + \underline{\mathit{82}}) - 126 = (A_{cl} - 44)$ u (*symmetric fission*) (20)

and eq.(12) is no more valid.

Fig.3 shows how the transfer of _82_ valence nucleons from the $A_H$= 126 nucleon core of the nascent heavy fragment to the primordial cluster, $^{50}Ar$, of $^{258}Fm$ fills not only the $A_L$= 82 nucleon shell of the light fragment but also its $A_L$ = 126 nucleon shell. Thus the only allowed mass-values, for light as well as for heavy $^{258}Fm$ fragments, are now: 126,127,128,129, 130,131 and 132; its means that only four fragment-pairs can be formed in $^{258}Fm$ according to the nucleon- phase hypothesis; this theoretical width A = 6 u is included in the hatched area of "appreciable yield" of symmetric fission, determined by eqs. 8, 18 and 19, and shown in the right part of fig.3.

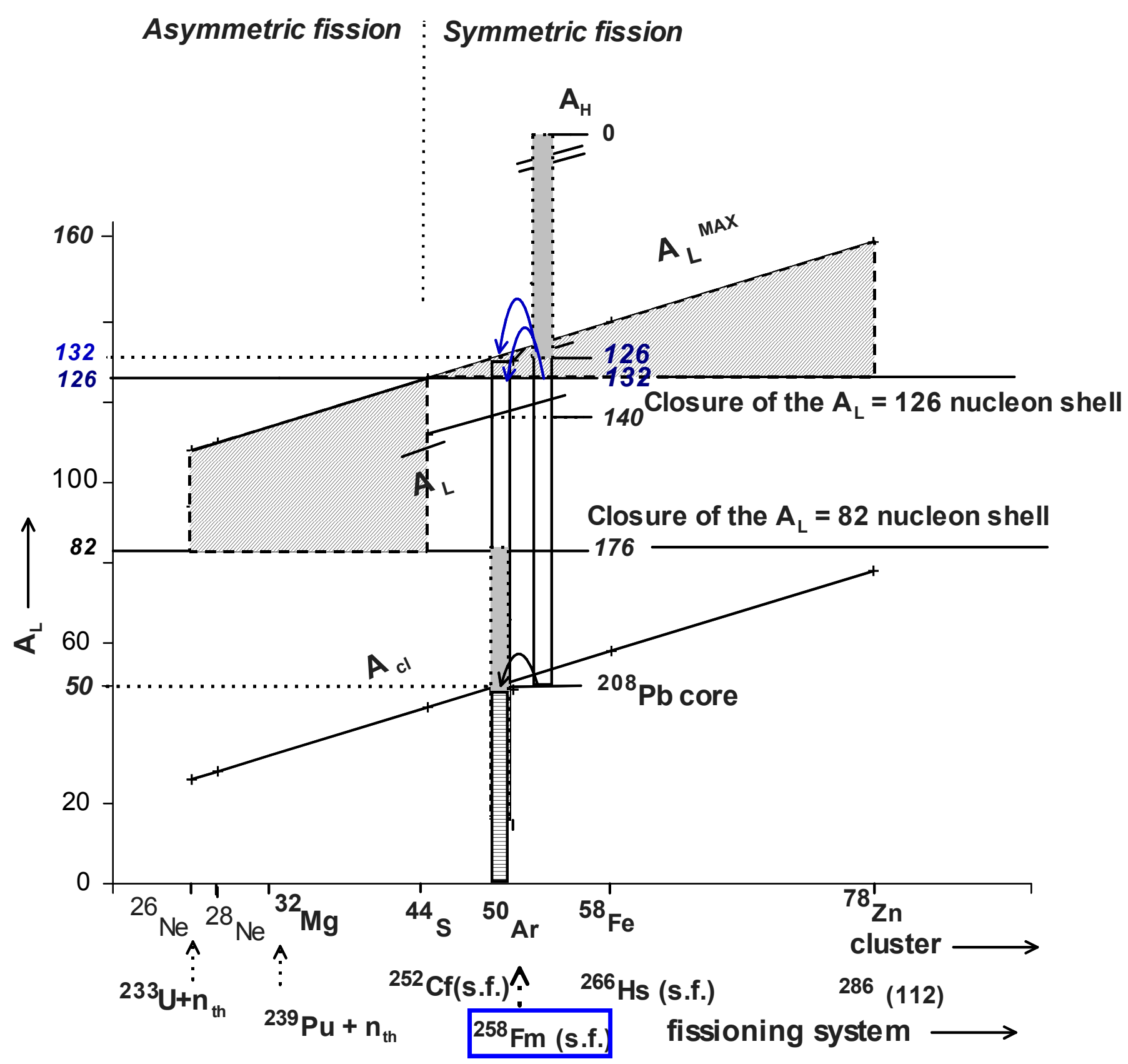

**Fig.3** Transfer <u>of 82 nucleons from core to cluster in the spontaneous fission of $^{258}Fm$ according to the nucleon-phase model.</u>
The region of appreciable yield for the symmetric fission of the heavy and superheavy nuclei is represented, up to $^{286}(112)$, by the hatched area on the right.

## 5-2 The phenomenon of barrier –free fission

The expected fragment pairs of $^{258}Fm$ are also the four Sn/Sn fragment pairs $^{126}Sn$-$^{132}Sn$, $^{127}Sn$-$^{131}Sn$, $^{128}Sn$-$^{130}Sn$ and $^{129}Sn$-$^{129}Sn$. Other charge splits can be neglected, since the uncertainty in Z is equal to ~ 1.6 u [1], i.e. very small, and since they are not as much energetically favored as the Sn/Sn splits.

However, G. Mouze found in 2005 [18] that only two of these Sn/Sn fragment pairs can be responsible for the peak at symmetry. These pairs,

$^{128}Sn$-$^{130}Sn$, Q = 253.794 MeV [2], (21)

$^{126}Sn$-$^{132}Sn$, Q = 252.895 MeV [2],

are so energy-rich, that their fission energy Q is greater than their own, duly corrected Coulomb barrier, namely

$B_C^{corr}$ = 252.191 MeV and (22)

$B_C^{corr}$ = 252.205 MeV , respectively.

The necessary "sphericity correction", applied to

$B_C$ (MeV) = $1.44\, Z_1 Z_2 / r_0 (A_1^{1/3} + A_2^{1/3})$, with $r_0 = 1.48$ (23)

has been estimated by a comparison of the fission properties of the three fermium nuclei, $^{256}Fm$, $^{257}Fm$ and $^{258}Fm$: $^{256}Fm$ does not fission symmetrically, whereas symmetric fission just becomes observable in $^{257}Fm$; the sphericity correction reaches + 11.49 MeV for the $^{128}Sn$-$^{130}Sn$ [18].

It is noteworthy that it is *the absence of any Coulomb barrier for the emission of the pairs $^{128}Sn$-$^{130}Sn$ and $^{126}Sn$-$^{132}Sn$ that explains their high yield.*

This situation has important consequences.

First, the full width at half maximum (f.- w.- h.- m.) of 8 u of the peak observed at A = 129 by D.C. Hoffman et al. [19] in $^{258}Fm$ can now be explained. To these two pairs only four favored masses correspond, namely 126,128,130 and 132. But if the uncertainty on mass of about 4 u [1] is taken into account, *the observed mass distribution has to be considered as resulting from the addition of four Gaussian curves*, each having a width of ~ 4 u, centered on these four masses: One verifies that a mass distribution constructed in this way has a f.- w.- h.- m. of 8 u , as found experimentally.

Secondly, the comparison of this f.- w.- h.- m. of 8 u of $^{258}Fm$ to the width $\Delta A$ of 6 u predicted by the nucleon- phase hypothesis (eq .(19)) for this nucleus suggests the following relation between the two widths:

f.- w.- h.- m. =( $\Delta A$ + ~ 2) u . (24)

This relation furnishes a useful, but approximate, estimation of the systematic *broadening,* caused by the uncertainty in mass of the fragments, of any symmetric fission.

Thirdly, due to the absence of any Coulomb barrier at symmetry, there is *no more reduction of yield in the vicinity of the theoretical limits* of the region of appreciable yield of symmetric fission: the full-width-at-half-maximum is really that given by eq. (24).

This situation is radically different from that encountered in asymmetric fission, e.g. in fig.1. The reason is that *asymmetric fission is a completely confined process.* There, no fragment pair has a Q- value greater than its own Coulomb barrier [6]; moreover, the difference $B_c$ – Q increases in going towards the limits, and causes a smaller tunnel-effect, as shown in fig.1.

But the most important consequence of the existence of barrier-free fission is that the bimodality property of symmetric fission can now be explained.

**5-3 The origin of the bimodality of symmetric fission.**

Let us recall that Hulet et al.[20] called “bimodal” the symmetric mode of fission of $^{258}$Fm, because they had found that its energy spectrum can be decomposed into two parts; more precisely, they attempted to analyze the total kinetic energy distribution by a Gaussian fitting and found that the mean energy of the low-energy component is equal to about 200 MeV – what is close to the value expected from the Viola systematics [21] –, whereas the mean energy of the higher component is considerably larger, namely about 233 MeV [20]. This analysis also revealed that the main contribution to the mass yield at symmetry comes from the high-energy events, whereas the low-energy events contribute to the broad pedestal of the mass distribution.

*The absence of any Coulomb barrier for emission of the most energy-rich fragments explains that their energy can be as great as the maximum energy released by the fission reaction*. Indeed, the energy spectrum of Hulet et al. extends up to ~ 260 MeV.

But, due to the uncertainty in mass [1], a large number of mass splits contribute to the peak at symmetry in the $^{258}$Fm case, and this fact explains the mean energy of only ~ 233 MeV of the high-energy component, since all these mass-splits cannot have the maximum energy of 253.8 MeV.

The low- energy component of the energy spectrum constitutes an indication that the asymmetric fission mode still survives and contributes to a low-intensity pedestal of the *mass* spectrum, which spreads over a great mass interval, also observed in $^{260}$Md by Ohtsuki et al.[12]; indeed, the low-energy component in the *energy* spectrum of $^{258}$Fm cannot be due to the sole still "confined " $^{127}$Sn-$^{131}$Sn and $^{129}$Sn-$^{129}$Sn pairs.

But this survival can also show itself in a *yield increase* in the region of symmetry, and consequently in an apparent *broadening* of the *"symmetric" mass distribution*, if excitation energy is furnished to the fissioning system; indeed, the "tunnel effect" then increases, because the "fission barriers $B_C^f$ (i) for a fission into a given pair i" , equal to $B_c$ (i) – Q (i), themselves decrease, being changed into $B_c$ (i) – [ Q (i) + $E_{exc}$ ]; and this occurs particularly around A = 132, because the Q(i) have their greatest values there.

**5-4 Superheavy nuclei and cluster-fission**

The symmetric mass distributions of very heavy nuclei, such as $^{266}$Hs [22], or $^{286}$(112)* [22] *have the widths predicted by the nucleon-phase hypothesis*, namely 16 u and 36 u, according to eq. (20), since the primordial clusters are, respectively $^{58}$Fe and $^{78}$Zn: This can be verified in the mass spectra reported by Itkis et al. [22].

Whereas the symmetric mass distribution of $^{258}$Fm is "*narrowly symmetric*" [19], that of $^{266}$Hs is "*broadly symmetric*" [23] and that of $^{286}$(112)* *double-humped* [24]. We have shown that this situation is the consequence of the role played by the phenomenon of barrier-free fission [23,24].

But a new phenomenon appears in these very heavy nuclei, that of *cluster-fission:* G. Mouze demonstrated in 2002 [25] that *the energy released by the primordial clusterization* of any fissioning system – an energy which is already equal to 59.49 MeV in $^{235}$U + $n_{th}$ and to 81.3 MeV in $^{239}$Pu + $n_{th}$ – increases rapidly as a function of $A_F$ and *becomes greater than the Coulomb barrier of the core-cluster pair if Z becomes greater than 111.* So cluster-fission can be considered as a new kind of barrier-free fission.

In low- energy fission, one principally observes peaks at 208 and at A = $A_F$ – 208, for example at A = 208 and at A = 78 in the case of $^{286}$(112)* [22], corresponding to the pair $^{208}Pb$-$^{78}Zn$. But at higher excitation energies, other peaks can be observed. All these structure are, more commonly, called "*quasi-fission*" structures [26].

The occurrence of cluster-fission structures such as those observed at A = 208 and A = $A_F$ – 208 in the mass distributions of very heavy nuclei may be considered as a further argument in favor of the ignition step of the fission process.

In conclusion of this Sect.5, it may be asserted that *the nucleon-phase hypothesis explains symmetric fission*, since it predicts the limits of the region of appreciable yield of the symmetric mass distributions.

But the phenomenon of barrier-free fission explains a number of facts, such as the "bimodality" of symmetric fission, the variation of the profile of the mass distributions in going from $^{258}Fm$ to $^{306}(122)^*$, and the considerable yield even at the limits of the distribution.

## 6. Conclusion.

The nucleon- phase hypothesis furnishes a *coherent* interpretation of a great number of experimental facts, which were unexplained up to now. Above all*, it explains both the asymmetric and symmetric mass distributions encountered in fission.*

Much work remains to be done in order to understand *the true nature* of this nucleon phase. It constitutes an *ephemeral new state of nuclear matter*, which should show itself in other processes than that of binary nuclear fission.